\documentclass[english,english,aps,prd,amsmath,amssymb,superscriptaddress]{revtex4}
\usepackage[T1]{fontenc}
\usepackage[latin9]{inputenc}
\usepackage{amsmath}
\usepackage{amssymb}
\usepackage{esint}

\makeatletter
\@ifundefined{textcolor}{}
{%
 \definecolor{BLACK}{gray}{0}
 \definecolor{WHITE}{gray}{1}
 \definecolor{RED}{rgb}{1,0,0}
 \definecolor{GREEN}{rgb}{0,1,0}
 \definecolor{BLUE}{rgb}{0,0,1}
 \definecolor{CYAN}{cmyk}{1,0,0,0}
 \definecolor{MAGENTA}{cmyk}{0,1,0,0}
 \definecolor{YELLOW}{cmyk}{0,0,1,0}
 }

\usepackage{slashed}
\usepackage{babel}
\usepackage{bbm}

\makeatother

\usepackage{babel}

\begin{document}

\title{Dark Energy from Quantum Uncertainty of Distant Clock}

\author{M.J.Luo}

\address{Department of Physics, Jiangsu University, Zhenjiang 212013, People's
Republic of China}
\begin{abstract}
The observed cosmic acceleration was attributed to an exotic dark
energy in the framework of classical general relativity. The dark
energy behaves very similar with vacuum energy in quantum mechanics.
However, once the quantum effects are seriously taken into account,
it predicts a completely wrong result and leads to a severe fine-tuning.
To solve the problem, the exact meaning of time in quantum mechanics
is reexamined. We abandon the standard interpretation of time in quantum
mechanics that time is just a global parameter, replace it by a quantum
dynamical variable playing the role of physical clock. We find that
synchronization of two spatially separated clocks can not be precisely
realized at quantum level. There is an intrinsic quantum uncertainty
of distant clock time, which implies an apparent vacuum energy fluctuation
and gives an observed dark energy density $\rho_{de}=\frac{6}{\pi}L_{P}^{-2}L_{H}^{-2}$
at tree level approximation, where $L_{P}$ and $L_{H}$ are the Planck
and Hubble scale cutoffs. The fraction of the dark energy is given
by $\Omega_{de}=\frac{2}{\pi}$, which does not evolve with the internal
clock time. The {}``dark energy'' as a quantum cosmic variance is
always seen comparable with the matter energy density by an observer
using the internal clock time. The corrected distance-redshift relation
of cosmic observations due to the distant clock effect are also discussed,
which again gives a redshift independent fraction $\Omega_{de}=\frac{2}{\pi}$.
The theory is consistent with current cosmic observations.
\end{abstract}
\maketitle

\section{Introduction}

The most important observational discovery of physics in the past
decade is the acceleration of the expanding universe \cite{Riess:1998cb,Perlmutter:1998np}.
In the standard model of cosmology based on the classical general
relativity, the mysterious driving force of the acceleration could
be simply attributed to a kind of energy unseen before, called dark
energy \cite{RevModPhys.75.559}. The observational studies of the
dark energy shows that it is (i) almost uniformly distributed, (ii)
very slowly varied with time and (iii) the equation of state is around
$w=-1$. 

If we only consider these three properties of the dark energy, it
behaves very similar with the vacuum energy we have already known
in quantum mechanics. However, if the quantum nature of vacuum is
seriously taken into account, it gives a disappointing wrong prediction
to its value \cite{RevModPhys.61.1}. The quantum mechanics predicts
that it is quartic divergent up to the ultraviolet cut-off. If the
validity of quantum mechanics is believed up to the Planck scale $10^{19}\mathrm{GeV}$,
the theory gives a very large prediction $(10^{19}\mathrm{GeV})^{4}$,
which is about $10^{120}$ times departure to the current observational
value $\rho_{de}\sim(10^{-11}\mathrm{GeV})^{4}$.

Compared with the small bare value, the large result would need to
be cancelled almost, but not exactly. It seems almost impossible to
explain the observed dark energy within the framework of conventional
quantum mechanics unless the theory is severely fine-tuned. The shortcoming
of the vacuum energy explanation gives room to other attempts to resolve
the problem, such as many phenomenological scalar fields dark energy
models \cite{2006IJMPD..15.1753C}, but unfortunately they are also
restricted in classical or semi-classical framework. These kinds of
models can also reproduce the above three properties and a correct
energy density within current range of observations, by carefully
tuning its kinetic term and classical potential to a specific shape.
In fact, even any behavior e.g. the time evolution of dark energy
and the equation of state around $w=-1$ can be engineered. Actually,
without quantum mechanics, a very small cosmological constant, phenomenologically,
also poses no problem. So the real question of dark energy in fact
concerns the inconsistent predictions between quantum mechanics and
general relativity.

The dark energy problem is a crisis deeply rooted in the foundation
of physics. It is known that the vacuum energy corrections to the
particle mass does gravitate \cite{2006hep.th....3249P}, and hence
there is by now no experimental evidence showing any violation of
the equivalence principle. If we trust the equivalence principle,
all energies gravitate, why we do not feel the large amount of quantum
vacuum energies by their gravitational effect, that is the first part
of the problem. It is an obvious contradiction between quantum mechanics
and the equivalence principle. If any mechanisms prohibit their gravitational
effects, why it seems that the quantum vacuum leaves a small remnant
gravitational effect which drives the cosmic acceleration, which is
the second part of the problem. Current observations bring forward
the third part of the problem: if the dark energy is a constant vacuum
energy, it is comparable with the matter energy density only in a
particular epoch, since the matter energy density is diluted as the
universe expanding, why the current observed vacuum energy is comparable
to the matter energy density or critical energy density now, which
is known as the coincidence problem or {}``why now'' problem \cite{2008GReGr..40..607B}. 

It would be a {}``mission impossible'' to solve these three aspects
of the problem, if our arguments are built upon the two foundations
mentioned: (i) standard quantum mechanics and (ii) the equivalence
principle of the general relativity. Remind that these two basis by
now still have not reconciled with each other, preventing a consistent
theory of quantum gravity, so it becomes more or less understandable
that these two theories would not give a consistent prediction to
the observed cosmic dark energy. The observed dark energy is likely
an experimental evidence for the confliction between these two theories.
As a general believe, the difficulty of reconciling the quantum mechanics
and the general relativity is deeply rooted in the very different
treatment of the concept of time \cite{rovelli2004quantum}.

\section{Quantum Uncertainty of Distant Clock}

In the quantum mechanics, time as a global parameter is independent
with where the clocks are placed on a space-like hypersurface. But
this statement is not true in all rigor when the quantum nature of
clocks is taken into account. In the spirit of relativity, time must
be operationally defined by a physical clock field $T(x)$ describing
the readings of e.g. a pointer's position of the clock, where $x=(x_{0},x_{1},x_{2},x_{3})\in\mathbb{R}_{4}$
are external space-time point parameter of the clock field in Euclidean
metric. The clock reading $T(x)$ is an internal time measured by
a local observer, while the external parameter $x$ can only be measured
by an external classical observer outside the universe. The physical
clock $T(x)$ is assumed to be a real scalar field, and satisfies
a zero-mass free field action,\begin{equation}
S_{T}=\int d^{4}x\frac{1}{2}\left(\partial_{x}T\right)^{2}.\label{eq:clock-action}\end{equation}

Now considering a thought experiment comparing the quantum states
of two spatially separated quantum clocks. The two quantum states
of the clocks placed at $x$ and $y$ here are described by states
$\left|T(x)\right\rangle $ and $\left|T(y)\right\rangle $. If the
norm of the inner product of these two quantum states equal to 1,
then these two states are identical, says, these two quantum clocks
are completely synchronized. The inner product is easy to calculate
according to the clock's action Eq.(\ref{eq:clock-action}), when
the space-like interval $\left|x-y\right|$ is considerable, we find
the asymptotic correlation between the clocks \begin{equation}
\left\langle T(x)|T(y)\right\rangle \sim\frac{1}{4\pi^{2}\left|x-y\right|^{2}},\label{eq:clock-correlation}\end{equation}
which decays with the distance between two clocks. For there is no
prior reason to tell us that whether the {}``same'' clocks spatially
separated are precisely synchronized, the gradual decorrelation exhibits
that the synchronization between two quantum clocks can not be precisely
realized. If we consider the clock at $y$ is standard (zero-uncertainty),
then the same clock at $x$ is uncertain. In a homogeneous, isotropic,
flat and empty space, considering a standard clock with reading $T(y)$
is transported from place $y$ to $x$, then the wavefunction that
one finds the clock at the distant place $x$ with reading $T(x)$
is given by \begin{align}
\int_{T(y)}^{T(x)}\mathcal{D}Te^{-S_{T}} & =\frac{V_{\mathbb{R}_{3}}^{2}}{4\pi^{2}\left|x-y\right|^{2}}e^{-2V_{\mathbb{R}_{3}}\frac{\left[T(x)-T(y)\right]^{2}}{\left|x-y\right|}}=\frac{1}{\sigma^{4}(2\pi)^{2}}e^{-\frac{4\left[T(x)-T(y)\right]^{2}}{2\sigma^{2}}},\label{eq:clock-correlation-2}\end{align}
where $\int\mathcal{D}T$ is the Feynman's path integral of the physical
clock. The width $\sigma^{2}$ of the wavefunction describes the uncertainty
of the reading $T(x)$ of the distant clock at $x$ with respect to
the standard clock at $y$, which is given by \begin{equation}
\sigma^{2}=\left\langle \delta T^{2}\right\rangle =\frac{1}{V_{\mathbb{R}_{3}}}\left|x-y\right|,\label{eq:clock-uncertainty}\end{equation}
where $V_{\mathbb{R}_{3}}$ is the 3-volume infrared cut-off. Therefore,
the simultaneity defined by physical clock $\left\langle T\right\rangle =\mathrm{constant}$
has an intrinsic quantum uncertainty increasing with the spatial interval
$\left\langle \delta T^{2}\right\rangle \propto\left|x-y\right|$.
Since the infrared cut-off 3-volume $V_{\mathbb{R}_{3}}$ here is
considered to be the cosmic scale but not infinity, the uncertainty
of simultaneity is not zero. It is a so small number that it can be
ignored in our ordinary observation, while it is considerable and
important when the spatial interval is at cosmic scale. By dimensional
consideration, the distant simultaneity uncertainty can be written
as\begin{equation}
\left\langle \delta t^{2}\right\rangle \sim L_{H}^{-3}L_{P}^{4}\left|x-y\right|,\label{eq:time-uncertainty}\end{equation}
where $L_{H}\sim V_{\mathbb{R}_{3}}^{1/3}$ and $L_{P}$ are the infrared
and ultraviolet cut-offs chosen as the Hubble and Planck scale. The
formula provides a universal limit to distant time measurement. In
general, if we consider the time is measured by a quantum physical
clock, but a global parameter, an intrinsic quantum uncertainty of
distant simultaneity is inevitable. It is worth emphasizing: (i) the
effect is different from the time dilation, it does not change the
central value $\left\langle t\right\rangle $ of the distant time,
it only makes the time fuzzy with a non-vanishing $\left\langle \delta t^{2}\right\rangle $.
(ii) Different from those time effects predicted from relativity,
in which time are different in different reference frames or in a
curved space, here, the effect even happens in one reference frame
and/or in a flat space. This quantum effect that a distant clock must
be uncertain provides a new explanation to the dark energy.

\section{Dynamical System Under Physical Clock}

To study the impact of the physical clock to a dynamical universe
system evolving with it, we consider that a whole system is defined
by including a clock field $S_{T}[T(x)]$ and the rest of the (to-be-measured)
universe $S_{U}[\varphi(x)]$ sharing the external parameter $x$.
These two systems are assumed independent and do not interact with
each other, while the time evolution of the rest of the universe $S_{U}$
is with respect to the clock field. So the action of the whole system
is separable \cite{rovelli1990quantum,PhysRevD.27.2885,Moreva:2013ska}
\begin{equation}
S=S_{U}+S_{T}.\end{equation}
Before studying the system, let us first briefly proof that the system
$S$ is semi-classically equivalent to the to-be-measured system $S_{U}$
where the conventional parameter time is used. Without loss of generality,
considering the to-be-measured system is a (one parameter) mechanical
system $S_{U}[\varphi(\tau)]=\int d\tau\frac{1}{2}\left(\partial_{\tau}\varphi\right)^{2}-V[\varphi]$,
and the physical clock is $S_{T}=\int d\tau\frac{1}{2}\left(\partial_{\tau}T\right)^{2}$,
then the partition function of the whole system is\begin{equation}
Z=\int\mathcal{D}\varphi\mathcal{D}Te^{-(S_{U}+S_{T})}.\end{equation}
The functional integral $\int\mathcal{D}T$ of physical clock can
be calculated by the mean field approximation, \begin{equation}
Z\overset{MF}{\approx}\int\mathcal{D}\varphi e^{-S_{eff}}.\end{equation}
Up to an unimportant constant, the effective action could be written
as \begin{equation}
S_{eff}\left[\varphi,\frac{\delta\varphi}{\delta T}\right]=\int dT\frac{1}{2}\mathcal{M}\left(\frac{\delta\varphi}{\delta T}\right)^{2}-V[\varphi]+\mathrm{constant},\end{equation}
in which $\mathcal{M}=\left\langle \left\Vert \frac{\partial\tau}{\partial T}\right\Vert \left(\frac{\partial T}{\partial\tau}\right)^{2}\right\rangle _{MF}$
is a constant depending on the integration constant of the mean field
value of $T(\tau)$. It is easy to see that the mean field value of
$T(\tau)$ is a monotonically increasing function of $\tau$, in this
sense, the quantum clock becomes classical. The effective action now
reproduces the classical structure of action $S_{U}$, only formally,
the functional derivative with respect to the clock time $T(\tau)$
replaces the conventional derivative with respect to the parameter
time $\tau$. 

The one-parameter proof can be generalized to a multi-parameter case,
in which not only time but also spatial coordinates are measured by
physical instruments. The multi-parameter case that puts the time
and space on an equal footing is equivalent to generalize the idea
of quantum clock to a quantum reference frame \cite{2014NuPhB.884..344L}. 

Generally speaking, the system $S=S_{U}+S_{T}$ corresponds to a system
satisfying a timeless Wheeler-DeWitt equation, while the system $S_{eff}$
corresponds to an emergent effective system (from $S$) satisfying
the Schrodinger equation in which external parameter time is used.
It is worth stressing that the theory $S$ and $S_{U}$ are equivalent
at semi-classical level, but they are different at quantum level.
The rest of the paper is based on the system $S=S_{U}+S_{T}$.

\subsection{Zero-Point Energy}

Since the notion of time now is changed, the notion of energy changes
accordingly. Energy is defined as a conserved quantity under the time
shift, and hence formally, the conventional derivative in the energy
definition $E\sim\frac{\partial}{\partial t}$ is replaced by a functional
derivative $E\sim\frac{\delta}{\delta T}$. Note that the action $S_{T}$
is quadratic in $T$, and $S_{U}$ does not explicitly contain $T$,
so the vacuum energy of the system is \begin{equation}
\left\langle E\right\rangle =-\frac{\delta\ln Z}{\delta T}\approx\frac{\delta S}{\delta T}=0.\end{equation}

This result means that the zero-point vacuum energy of the whole system
$S$ is vanished under the physical time $T$, which explains the
first part of the problem. The physical reason for that the zero-point
energy $\frac{1}{2}\sum_{k}\hbar\omega_{k}$ does not appear is transparent,
because here time is the internal field $T(x)$ undergoing quantum
fluctuation but an external parameter time $x_{0}$, the zero-point
energy can not be seen when the observer is holding a physical clock
that is also quantum fluctuating.

\subsection{Vacuum Energy Fluctuations}

That is not to say the vacuum is trivial, according to the uncertainty
principle, an apparent energy variance emerges out of the void related
to the intrinsic time uncertainty Eq.(\ref{eq:time-uncertainty}),
i.e. $\left\langle \delta E^{2}\right\rangle =\left\langle E^{2}\right\rangle -\left\langle E\right\rangle ^{2}=\left\langle E^{2}\right\rangle =\frac{\delta^{2}S}{\delta T^{2}}\neq0$.
The further the distance, the more uncertain the time, and the larger
the energy variance out of the void. The vacuum energy fluctuation
in a 4-volume element can be given by\begin{align}
\left\langle \delta E(x)\delta E(0)\right\rangle d^{4}x & =-\frac{\delta^{2}\ln Z}{\delta T(x)\delta T(0)}d^{4}x\nonumber \\
 & \approx\frac{\delta^{2}S}{\delta T(x)\delta T(0)}d^{4}x=\partial_{x}^{2}\delta^{4}(x)d^{4}x.\end{align}
At tree level approximation, we have approximately used $\ln Z\approx-S$,
so the leading result is expressed in terms of a widthless Dirac delta
function, while it actually has a non-zero width. This calculation
can be performed by first rewrite the Dirac delta distribution as
a limit of the Gaussian distribution, i.e. $\delta(x)=\lim_{a\rightarrow0}\frac{1}{a\sqrt{\pi}}e^{-\frac{x^{2}}{a^{2}}}$,
doing the derivatives and finally taking the zero width limit of the
Gaussian distribution back to the Dirac delta distribution,\begin{align}
\left\langle \delta E(x)\delta E(0)\right\rangle d^{4}x & =\lim_{a\rightarrow0}\partial_{x}^{2}\left(\frac{1}{a\sqrt{\pi}}e^{-\frac{x^{2}}{a^{2}}}\right)^{4}d^{4}x\nonumber \\
 & =64a^{-4}\left|x-0\right|^{2}\delta^{4}(x)d^{4}x.\label{eq:energy-variation}\end{align}
The width of the Gaussian distribution $a$ is an ultraviolet cut-off,
the most natural choice is the Planck length $a=L_{P}$. If the distance
$\left|x-0\right|$ is large, the energy fluctuation becomes considerable
when it is at cosmic scale.

To regulate the result, an infrared cut-off is required, a natural
choice is the Hubble length $\left|x-0\right|=L_{H}$, as the largest
distance we could see, i.e. cosmic horizon. Therefore, when we fix
the radius $\left|x-0\right|=L_{H}$ and integrate over $x$, then
the total energy fluctuation of the vacuum in the Hubble scale volume
is obtained\begin{equation}
\left\langle \delta E^{2}\right\rangle =64\int d^{4}xL_{P}^{-4}L_{H}^{2}\delta^{4}(x)=64L_{P}^{-4}L_{H}^{2}.\end{equation}
Then an averaged vacuum energy density (averaged in the 3-ball with
fixed radius $\left|x-0\right|=L_{H}$) due to the total vacuum energy
fluctuation is predicted as\begin{equation}
\rho_{de}=\frac{\sqrt{\left\langle \delta E^{2}\right\rangle }}{\frac{4\pi}{3}L_{H}^{3}}=\frac{6}{\pi}L_{P}^{-2}L_{H}^{-2},\label{eq:de-density}\end{equation}
and \begin{equation}
\Omega_{de}=\frac{\rho_{de}}{\rho_{c}}=\frac{2}{\pi}\approx0.64,\label{eq:fraction}\end{equation}
where $\rho_{c}=\frac{3H^{2}}{8\pi G}$ is the critical density, $H=L_{H}^{-1}$
is the Hubble's constant, $8\pi G=L_{P}^{2}$ is the Newton's gravitational
constant, and $\Omega_{de}$ is the fraction of the effective vacuum
energy. The leading order predicted $\Omega_{de}$ is a little lower
than the current best fit from the data of Planck satellite \cite{2013arXiv1303.5062P},
but still within the allowed range. This result explains the second
part of the problem.

There are several important remarks of this result to emphasize. (i)
We have considered the question: what a vacuum energy fluctuation
is seen by a distant observer in a homogeneous, isotropic and empty
flat space, when the time is defined by a physical clock field $T(x)$.
(ii) The coordinates $x$ in the action Eq.(\ref{eq:clock-action})
are just external parameters which can only be seen by an external
classical observer outside the universe. The reason we only pick up
{}``time'' treating quantum mechanically and the spacetime coordinates
treating as the external parameter is for simplicity, a more rigor
and general quantum mechanical treatment is to put the space and time
on an equal footing (quantum reference frame \cite{2014NuPhB.884..344L}),
which does not dramatically change the result when we only focus on
energy and time. (iii) The value of the results Eq.(\ref{eq:de-density}
and \ref{eq:fraction}) gives correct order, in fact, they indeed
depend on the precise nature of the cut-offs, and at present, the
numerical factors of the cut-offs are chosen as the most natural ones.

\subsection{The Coincidence Problem}

Since the action $S_{T}$ is quadratic in $T$, the higher order ($>2$)
functional derivative with respect to clock time $T$ are all vanished,
i.e. $\frac{\delta\left\langle \delta E^{2}\right\rangle }{\delta T}=\frac{\delta^{3}S}{\delta T^{3}}=0$.
As a result, the vacuum energy fluctuation does not evolve with the
clock time, thus leading to the fraction $\Omega_{de}$ does not vary
with this clock time. It is a constant and is {}``always'' comparable
with the critical density. And as a vacuum energy, it is uniform and
a constant, moreover, its equation of state strictly equal to -1 and
does not vary with time either.

Note that, because in our framework the time is a local internal observable,
so those old notions of evolution in the standard cosmology must be
carefully reconsidered. The exact meaning of the time evolution of
any quantities is that their functional derivative with respect to
the clock time $T$ is non-vanished. However, for example, the Hubble
parameter $H(t)$, the fraction $\Omega_{i}(t)$ and the equation
of state $w(t)$ as functions of parameter time only seen by an external
classical observer have no physical meaning in our setting. So the
infrared cut-off, the Hubble constant $H$ and/or Hubble length $L_{H}$
in Eq.(\ref{eq:de-density}), in this sense, is really a constant.

As a consequence, we have had two statements: (i) in the standard
cosmology, the matter density evolves with time, while the dark energy
remains a constant; (ii) in this theory without coincidence problem,
the matter density is {}``always'' comparable with the dark energy
density. It seems that they contradict each other, how could these
two statements are both true, please do not immediately make an arbitrary
judgment that it must be wrong. The key is again the notion of time,
and the expansion of the universe is relative but absolute.

The view of a local internal observer is very different from that
of an absolute external observer who feels a gradually diluting matter
component under the cosmic expansion. If the universe is spatially
flat $\Omega_{K}=0$, the matter density is always approximately $\Omega_{M}\approx1-\Omega_{de}$
seen by an internal observer at any epoch. In contrary to what one
may think, here the matter density does not change under the local
internal clock. Considering the universe is divided into two parts,
one is a finite regime A in which an observer lives, and the regime
B is the rest of the universe. The notion {}``now'' in principle
is a limit of regime A shrinking to infinitely small, but in practice
the regime can be considered finite, i.e. the notion {}``near now''
or {}``a near epoch'' used above. The change in the regime B is
defined relative to the clock in regime A which is external. While
the change in the regime A is relative to the clock also in regime
A which is internal. The consequence is that the internal observer
does not see expansion of regime A with respect to the internal clock.
Because the internal observer always lives in the regime A ({}``near
now'' regime), although he/she as an external observer can see changes
in regime B, there seems an almost static matter density in the regime
A, since his/her rulers and clocks are expanding correspondingly,
that is the reason the internal observer always see the matter density
does not vary with time and always comparable with the apparent {}``dark
energy''. In this sense, the Hubble volume is seen unchanged with
local internal time, and hence the Hubble length as the infrared cutoff
always remains $L_{P}=H^{-1}$. In the standard external observer's
interpretation, it is a problem of coincidence, but in a local internal
observer's view, the densities do not vary with their clocks, and
the coincident redshift $z_{c}$ is always relatively small.

It is worth emphasizing that {}``always comparable'' does not mean
these two as real components of the universe would be scaled in the
same way under expansion seen by an external observer, since it is
impossible to be consistent with many observations such as the galaxies
formation and the growth of large scale structure. In certain sense,
the evolution of the observable universe gives place to the evolution/scaling
with redshift. The cosmic acceleration in fact is an apparent quantum
cosmic variance, and we are not living in a special epoch, whenever
an (internal) observation is performed, the mirage {}``dark energy''
is always seen being of the order of the matter density. What the
internal observer sees is very different from that of the standard
external observer. In this sense, the resolution of the coincidence
problem is not dynamical, the key is again the notion of time.

\section{Distance-Redshift Relation}

The notion of time is a key to the dark energy problem, this can be
seen also from analyzing what we really measure in those dark energy
observations \cite{Durrer:2011gq}. Up to date, the measurement indications
for the existence of dark energy (e.g. the supernovae Ia and Cosmic
Microwave Background (CMB)) come from the distance measurements $D$
and their relation to the redshift $z$. In fact, we have not measured
the dark energy and its equation of state directly. The two observables
$(z,D)$ are independently measured. The distance of supernovae is
determined by observing the {}``luminosity distance'', and the CMB
is again by the {}``angular diameter distance'' measurement on the
last scattering surface. The redshift of the supernovae and CMB relate
to the frequencies or time measurement of distant objects. Most of
the data satisfying the Hubble's law, which states a linear dependence
between the distance and redshift, is at low redshift regime. It is
the high redshift observations that detect a distance which is significantly
larger than the expected value in a flat matter dominanted or curvature
dominated universe with the same Hubble constant. The unusual $D(z)$
relation at high redshift then infers the existence of dark energy
by assuming the validity of general relativity. Until now there is
no other test to tell us whether the dark energy is truly a new component
of universe or simply a misunderstanding of the distant measurements,
especially at high redshift (far off distance) regime, for example
the cosmic scale distant frequency or time measurement. In fact, there
is no experimental basis to state that a distant measured frequency
or distant clock is exactly the same as the native ones.

Let us assume the time uncertainty previously considered in flat space
is still (at least approximately) correct in the Hubble's expanding
universe. At small redshift $z\equiv a_{0}/a-1$, the distance-redshift
relation $D(z)$ is given by \begin{equation}
H_{0}D=z+\frac{1}{2}z^{2}+...,\end{equation}
where $H_{0}$ is the Hubble's constant at $z=0$, $D$ is the luminosity
distance. Since the distant frequency or redshift measurement has
been reconsidered, such effect will give a modification to the distance-redshift
relation. The distant time uncertainty does not change the central
value of the spectral line or redshift, only broadens it and gives
a non-vanishing variance $\left\langle \delta z^{2}\right\rangle \neq0$.
As a consequence, the distance-redshift relation $D(z)$ is modified
at the order $\mathcal{O}(z^{2})$ by an extra positive contribution\begin{equation}
H_{0}D=\left\langle z\right\rangle +\frac{1}{2}\left(\left\langle z\right\rangle ^{2}+\left\langle \delta z^{2}\right\rangle \right)+...,\label{eq:distance-redshift-1}\end{equation}
in which we have used $\left\langle z^{2}\right\rangle =\left\langle z\right\rangle ^{2}+\left\langle \delta z^{2}\right\rangle $.
It is the extra positive contribution coming from the distant time/simultaneity
uncertainty makes the effective {}``dark energy'' behave repulsive.

Now we calculate the redshift variance $\langle\delta z^{2}\rangle$
from the uncertainty of distant clock given by Eq.(\ref{eq:clock-uncertainty}).
If we work in Minkovski space, i.e. $(ix_{0},x_{1},x_{2},x_{3})$,
the path integral Eq.(\ref{eq:clock-correlation-2}) becomes complex,
when the distance $\left|x-y\right|$ (and corresponding $V_{\mathbb{R}_{3}}\sim\left|x-y\right|_{max}^{3}$)
tends to infinity, we have a limit \begin{align}
\lim_{\left|x-y\right|\rightarrow\infty}\sqrt{\frac{V_{\mathbb{R}_{3}}}{2\pi\left|x-y\right|}}e^{\frac{iV_{\mathbb{R}_{3}}\left[T(x)-T(y)\right]^{2}}{2\left|x-y\right|}} & =\lim_{\left|x-y\right|\rightarrow\infty}\frac{1}{\sqrt{2\pi}\left(\delta T\right)}e^{\frac{i\left[T(x)-T(y)\right]^{2}}{2\left(\delta T\right)^{2}}}=e^{i\frac{\pi}{4}}\delta[T(x)-T(y)],\end{align}
where $\delta[T(x)-T(y)]$ is the Dirac delta function. Note that
the phase factor of the formula tends to a constant $\pi/4$ in this
limit, and hence the redshift variance over redshift squared tends
to $2/\pi$,\begin{equation}
\lim_{\left|x-y\right|\rightarrow\infty}\frac{(\delta T)^{2}}{[T(x)-T(y)]^{2}}=\lim_{z\rightarrow\infty}\frac{\langle\delta z^{2}\rangle}{\langle z\rangle^{2}}=\frac{2}{\pi},\end{equation}
in which we have used $\langle\delta z^{2}\rangle=\frac{\langle\delta T^{2}\rangle}{\langle T(x)\rangle^{2}}$
and $\langle z\rangle=\frac{\langle T(y)-T(x)\rangle}{\langle T(x)\rangle}$.

The linear relationship between the distant spectral line width and
redshift (at high redshift) is an important prediction of the idea
which can be tested by observations. Then we have a modified distance-redshift
relation, the Eq.(\ref{eq:distance-redshift-1}) becomes\begin{equation}
H_{0}D=\left\langle z\right\rangle +\frac{1}{2}\left(1-q_{0}\right)\left\langle z\right\rangle ^{2}+...=\left\langle z\right\rangle +\frac{1}{2}\left(1+\frac{2}{\pi}\right)\left\langle z\right\rangle ^{2}+...\end{equation}
Therefore, in a flat universe without ordinary matter (pressureless
matter and radiation), the uncertainty of distant clock induces a
redshift independent deceleration parameter $q_{0}=-\frac{2}{\pi}<0$,
which makes the flat empty universe seem accelerating and being dominated
by {}``dark energy'' with fraction $\Omega_{de}=\frac{2}{\pi}$.
This result deduced from an independent calculation agrees with Eq.(\ref{eq:fraction}).

In the flat universe $\Omega_{K}=0$, if the contribution of ordinary
pressureless matter is taken into account, which evolves as $\Omega_{M}(1+z)^{3}$
with the redshift $z$, then we have\begin{equation}
H_{0}D=\left\langle z\right\rangle +\frac{1}{2}\left(1-\Omega_{M}(1+z)^{3}+\Omega_{de}\right)\left\langle z\right\rangle ^{2}+...\end{equation}
So the universe is always seen become accelerating at a relatively
small redshift, at $-q_{0}=-\Omega_{M}(1+z)^{3}+\Omega_{de}=0$, i.e.
$z_{c}\approx0.3$ with respect to current epoch.

\section{Conclusions}

Finally, let us summarize the paper. In this paper, we retain the
equivalence principle and abandon the standard interpretation of parameter
time in quantum mechanics. The quantum spatial evolution makes the
physical clock field fuzzy as the distance increases, leading to a
quantum uncertainty of distant clock time. The idea of reinterpretation
of time solves the dark energy problem. This theory tells us that
the observed dark energy is a quantum effect connected to the quantum
uncertainty of spatially separated clocks. The apparent vacuum energy
fluctuation is inevitable if we use the physical clock redefining
the time, and the result fits the observation well. This framework
requires a modification of the standard quantum mechanics. Although
the global parameters of quantum mechanics are necessary by its intrinsic
structure, there is no prior reason to interpret them as time, time
here is what we read from a physical clock that needs to be described
quantum mechanically. The modified quantum framework requires a relational
interpretation in terms of entangled state which is more natural than
its standard absolute interpretation, since not only the to-be-measured
system but also the measuring instruments such as the clock are both
needed to be treated by quantum mechanics. In this sense, the Wheeler-DeWitt
equation plays a more fundamental role than the emerged Schrodinger
equation. And most importantly, this idea provides a touchstone to
the longstanding difficulty of reconciliation of the inconsistency
between general relativity and quantum mechanics.

\section*{Acknowledgments}

This work was supported in part by the National Science Foundation
of China (NSFC) under Grant No.11205149.

\bibliographystyle{apsrev}

\end{document}